\documentclass{egpubl}
\usepackage{eurovis2021}

\SpecialIssuePaper         

\usepackage[T1]{fontenc}
\usepackage{dfadobe}  

\usepackage{cite}  
\BibtexOrBiblatex
\electronicVersion
\PrintedOrElectronic
\ifpdf \usepackage[pdftex]{graphicx} \pdfcompresslevel=9
\else \usepackage[dvips]{graphicx} \fi

\usepackage{egweblnk}

\graphicspath{{Figures/}{./}}

\usepackage{amsfonts}
\usepackage{amsmath}
\usepackage{amssymb}

\usepackage{color}
\usepackage{enumitem}

\usepackage{scalerel,xparse}

\title[RAMPVIS: Visualisation Capabilities for Large-scale Emergency Responses]%
      {RAMPVIS: Towards a New Methodology for Developing Visualisation Capabilities for Large-scale Emergency Responses}

\author[M. Chen et al.]
{\parbox{\textwidth}{\centering%
  M. Chen$^{1}$\orcid{0000-0001-5320-5729},
  A. Abdul-Rahman$^{2}$\orcid{0000-0002-6257-876X},
  D. Archambault$^{3}$\orcid{0000-0003-4978-8479},
  J. Dykes $^{4}$\orcid{0000-0002-8096-5763},
  A. Slingsby$^{4}$\orcid{0000-0003-3941-553X},
  P. D. Ritsos$^{5}$\orcid{0000-0001-9308-3885},
  T. Torsney-Weir $^{3}$,
  C. Turkay $^{6}$\orcid{0000-0001-6788-251X},\\
  B. Bach $^{7}$,
  A. Brett $^{8}$,
  H. Fang $^{9}$\orcid{0000-0001-9365-7420},
  R. Jianu $^{4}$\orcid{0000-0002-5834-2658},
  S. Khan $^{10}$\orcid{0000-0002-6796-5670},
  R. S. Laramee $^{11}$\orcid{0000-0002-3874-6145},
  P. H. Nguyen $^{12}$\orcid{0000-0001-5643-0585},
  R. Reeve $^{13}$\orcid{0000-0003-2589-8091},\\
  J. C. Roberts $^{5}$\orcid{0000-0001-7718-3181},
  F. Vidal $^{5}$\orcid{0000-0002-2768-4524},
  Q. Wang $^{11}$,
  J. Wood $^{4}$\orcid{0000-0001-9270-247X}, and
  K. Xu $^{14}$\orcid{0000-0003-2242-5440}
  }
  \\
{\parbox{\textwidth}{\centering
  $^1$University of Oxford, \;
  $^2$King's College London, \;
  $^3$Swansea University, \;
  $^4$City, University of London, \;
  $^5$Bangor University, \;
  $^6$Warwick University,\\
  $^7$University of Edinburgh, \;
  $^8$UK Atomic Energy Authority, \;
  $^9$Loughborough University, \;
  $^{10}$Horus Security Consultancy Ltd., \;
  $^{11}$Nottingham University,\\
  $^{12}$Red Sift Ltd., \;
  $^{13}$University of Glasgow, \; and
  $^{14}$Middlesex University London, \quad United Kingdom
  } 
}
}

\begin{document}


\maketitle
\begin{abstract}
The effort for combating the COVID-19 pandemic around the world has resulted in a huge amount of data, e.g., from testing, contact tracing, modelling, treatment, vaccine trials, and more. In addition to numerous challenges in epidemiology, healthcare, biosciences, and social sciences, there has been an urgent need to develop and provide visualisation and visual analytics (VIS) capacities to support emergency responses under difficult operational conditions. In this paper, we report the experience of a group of VIS volunteers who have been working in a large research and development consortium and providing VIS support to various observational, analytical, model-developmental and disseminative tasks. In particular, we describe our approaches to the challenges that we have encountered in requirements analysis, data acquisition, visual design, software design, system development, team organisation, and resource planning. By reflecting on our experience, we propose a set of recommendations as the first step towards a methodology for developing and providing rapid VIS capacities to support emergency responses.


\end{abstract}

\section{Introduction}
\label{sec:Introduction}

Visualisation and visual analytics (abbreviated as VIS) has been used extensively in many mission-critical applications and healthcare applications.
Since the emergence of COVID-19, data visualisation has been widely visible in traditional and online media for disseminating information related to COVID-19.
Meanwhile what has not been obvious to the public is the fact that VIS techniques can be or have been used to help domain experts in combating COVID-19.
For example, mathematically, deriving an optimal model to forecast the contagion patterns of COVID-19 in different conditions (e.g., geographical, social, seasonal variation; different human intervention; etc.) is an intractable problem.
While modelling is an important tool for scientists to obtain a good understanding of COVID-19 and to assist decision-makers in exploring different intervention policies, it is necessary to provide modelling scientists and epidemiologists with a collection of VIS techniques that enable them to perform various observational, analytical, and model-developmental tasks rapidly and easily.

RAMP VIS \cite{RAMPVIS:2020:web} is a group of VIS volunteers, who answered a call to support the modelling scientists and epidemiologists in the Scottish COVID-19 Response Consortium (SCRC) \cite{SCRC:2020:web}.
The group was assembled as part of the rapid responses organised by the Royal Society (UK) \cite{RAMP:2020:web}.
As a volunteering operation in an emergency context, the VIS volunteers encountered many challenges. For example,
the time urgency demanded rapid development of usable VIS tools,
the travel restriction and the domain experts' heavy workload hampered in-depth requirements analysis, and parallel developments of pandemic models and data infrastructure entailed delays in accessing data to be visualised.
In addition, there was a shortage of skilled developers for designing and engineering a VIS system, and a fair amount of uncertainty in organising and scheduling volunteering resources.

The VIS volunteers made time urgency as their top priority, and were grouped into seven teams according to the available VIS expertise as well as different VIS needs in the SCRC modelling workflow. The grouping also enabled each team to progress independently in terms of requirements analysis, visual design, and system engineering. To our best knowledge, the experience of RAMP VIS volunteers is unprecedented in the VIS literature. In this paper, we describe our approaches to address the challenges in the overall RAMP VIS organisation as well as in the activities of individual teams. By reflecting on our experience, we propose a set of recommendations as the first step towards a methodology for developing and providing VIS capacities to support an emergency response. The main contribution of this work is thus the report of the RAMP VIS experience, our reflection, and a proposed methodology.
\section{Related Work}
\label{sec:RelatedWork}

In this section, we review the application of VIS to emergency response and healthcare and discuss the existing methodologies that may be used to develop VIS in such applications. 

\noindent\textbf{VIS for Emergency Responses.}
Emergency response has been a regular theme in VIS since 2005 \cite{Dusse2016}.
Kwan and Lee \cite{Kwan2005} incorporate geospatial visualisation into a real-time 3D emergency response system to support quick response to terrorist attacks.
Chittaro et al. \cite{Chittaro2006} introduce VU-Flow, a 3D visual environment that provides navigation guidance to users during emergency simulations.
Based on the data collected from large community disaster events (e.g., 9/11 and Hurricane Katrina), Campbell et al. \cite{Campbell2008} use visualisation-based interactive simulation for training emergency response teams.
Natarajan and Ganz \cite{Natarajan2009} introduce distributed visual analytics for managing emergency response between geographically dispersed users.
Waser et al. \cite{Waser2011a} incorporate visual designs into simulation-based investigation of flood disasters to recommend appropriate response strategies. Maciejewski et al. \cite{Maciejewski2011a} introduce PanViz, a VIS toolkit providing decision support for simulated pandemic scenarios.
Ribicic et al. \cite{Ribicic2012} develop a VIS interface to provide flood simulations to non-expert users. 
Konev et al. \cite{Konev2014} incorporate interactive visual designs into a simulation-based approach for flood protection planning.
Gelernter et al. \cite{Gelernter2018} provide visualisations to guide first responders at a crisis scene.
Visualising social media data is commonly used to support emergency responses
for situational awareness \cite{MacEachren2011}, resource allocation \cite{Jeitler2019}, critical infrastructure management \cite{Thom2016} and post-disaster analytics \cite{Hornbeck2019, Medoc2019,Nguyen2019a}.
Whitlock et al. \cite{Whitlock2019a} integrate VIS tools with mobile and immersive technologies to support critical operations during emergency response.

VIS plays a critical role in mission-critical applications such space missions. Abramyan et al. \cite{Abramyan2012} develop an immersive visualisation environment for controlling space robots remotely on the Earth. Edell and Wortman \cite{Edell2015} introduce advanced visualisations to assist the diagnosis of operational problems and failures for Van Allen Probes, a NASA space mission.

Furthermore, the IEEE Conference on Visual Analytics Science and Technology (VAST) host an international challenge workshop annually since 2006. Competition entries demonstrated novel VIS solutions for epidemic spread \cite{VAST2010MC2,VAST2011MC1}, illicit activities \cite{VAST2011MC3,VAST2018MC2,VAST2018MC3}, security streaming data \cite{VAST2016MC1,VAST2016MC2,VAST2016MC3} and natural disasters \cite{VAST2019}.

Unlike most of the prior work generally carried out in preparation for a future emergency, this work was conducted during the period of an emergency response to the COVID-19 pandemic. 

\noindent\textbf{VIS for Healthcare.}
The healthcare industry benefits from the adoption of Visualisation and Visual Analytics.
Rind et al. \cite{Rind2013} review VIS tools developed for the exploration of electronic health records.
Carroll et al. \cite{Carroll2014} review 88 articles on VIS tools for infectious disease.
Gotz and Borland \cite{Gotz2016} discuss challenges unique to the healthcare industry and the critical role that VIS plays in the domain.
McNabb and Laramee \cite{McNabb2017a} conduct an extensive survey of surveys including the adoption of VIS in the healthcare sector.
Preim and Lawonn \cite{Preim2020} survey the use of VIS for supporting decision making in the public health sector.

The VIS techniques used in healthcare often incorporate analytical techniques.
For example, \emph{Event Sequence Simplification} is used to reduce visual complexity of sequential clinical events \cite{Wongsuphasawat2011,Monroe2013,Gotz2014,Guo2019}, and
support a more efficient decision support process \cite{Augusto2005}.
\emph{Natural Language Processing} is used to extract textual data from raw clinical datasets \cite{Zhang2011,Glueck2017,Sultanum2019}.
\emph{Machine Learning} help automating the processing of clinical data and providing guidance to clinicians and researchers, including Active Learning \cite{Bernard2015}, Support Vector Machine \cite{Trivedi2018}, Topic Modelling \cite{Glueck2018} and Recurrent Neural Networks \cite{Kwon2019}.

The main difference between this work and the prior work is that we had to develop multiple VIS capacities rapidly to address different VIS needs concurrently.

\noindent\textbf{Methodologies for VIS Applications.} 
Design study methodology~\cite{12Sedlmair} builds on the nested model of design and validation~\cite{10Munzner} to provide guidelines, pitfalls, and a process that help visualisation researchers design systems in applied contexts. This methodology uses the metaphor of ``the Trenches'', which is where we found ourselves, and so, just as others have adapted its concepts and processes across varied settings~\cite{18Lam, 20Lite}, this paper reports our approach that adopts, adapts and sometimes contradicts established guidance in the context of rapid emergency response.

Workshops \cite{kerzner2019framework, knollBeliv2020} can speed up requirements gathering.
The five-design sheets methodology \cite{RobertsFDS_2016} can structure the sketching process.
Collaborative design methodologies \cite{losev_beliv_2020} can address issues due to travel restrictions, and assist rapid visualisation design processes~\cite{Dixit:2020:JAMIA}.
Some prior work in VIS advocates different software engineering methodologies~\cite{brooks_mmm_1975, pragmatic_programmer}.
The agile approach (Kanban boards, SCRUM, etc.) is particularly suitable for developing VIS systems in applications with changing characteristics of data, users, and tasks.
A recently-proposed method based on the cost-benefit analysis can potentially be used to discover shortcomings in a VIS workflow and explore potential solution systematically \cite{Chen:2019:CGF}.
We have drawn inspiration from these methodologies in our work.

\section{Formulating the RAMPVIS Approach}
\label{sec:Approach}

\emph{The Scottish COVID-19 Response Consortium} (SCRC) \cite{SCRC:2020:web} was established in April 2020 by researchers in three Scottish organisations in response to a call from the Royal Society for \emph{Rapid Assistance in Modelling the Pandemic} (RAMP) \cite{RAMP:2020:web}.
The goal of the Consortium was to develop a more robust and clearer understanding of potential medium- and long-term strategies for controlling the COVID-19 epidemic in Scotland and in the UK. The Consortium currently has over 150 members from 36 organizations.

\begin{figure}[t]
    \centering
    \begin{tabular}{@{}c@{\hspace{2mm}}c@{}}
        \includegraphics[height=33mm]{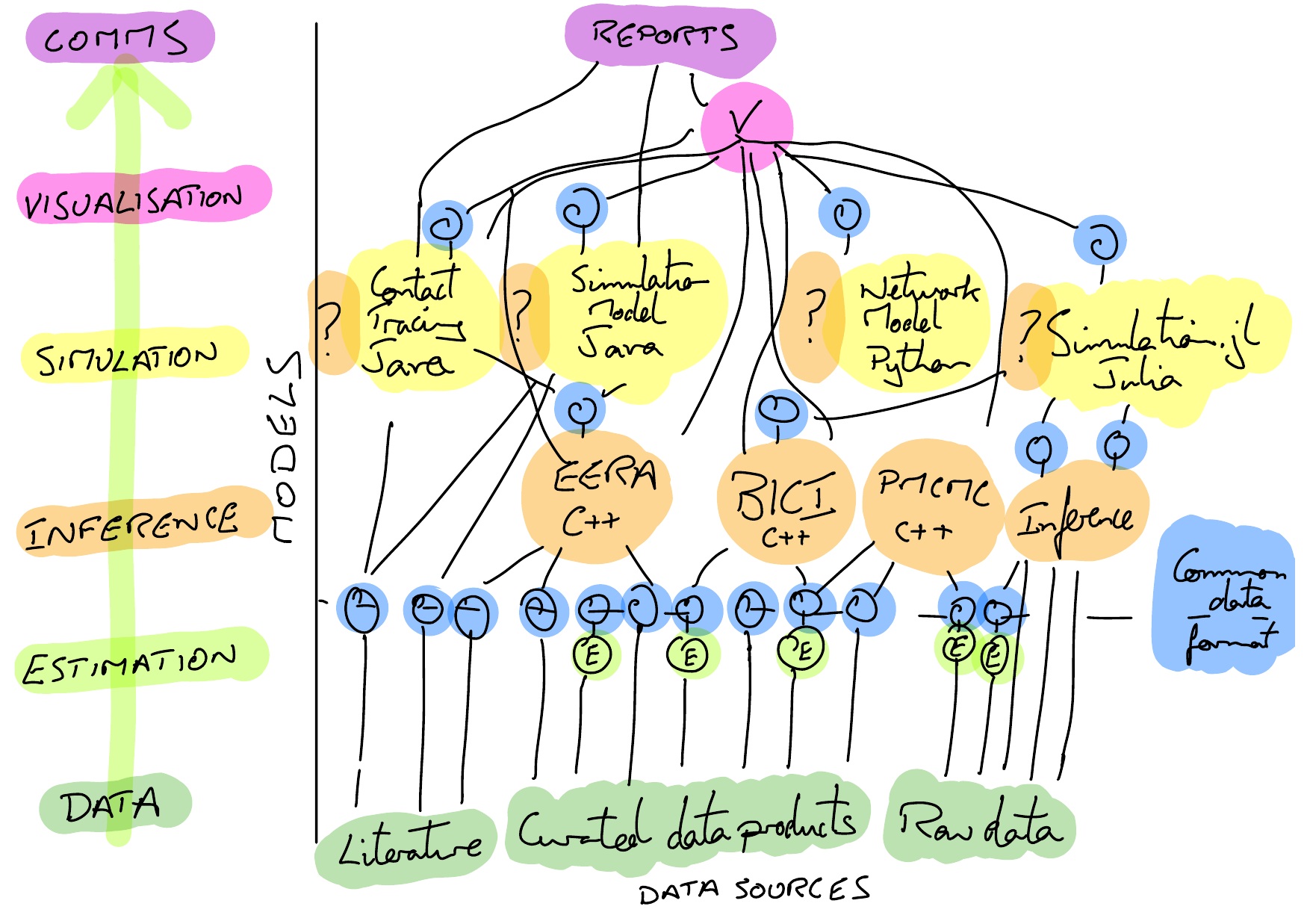} &
        \includegraphics[height=33mm]{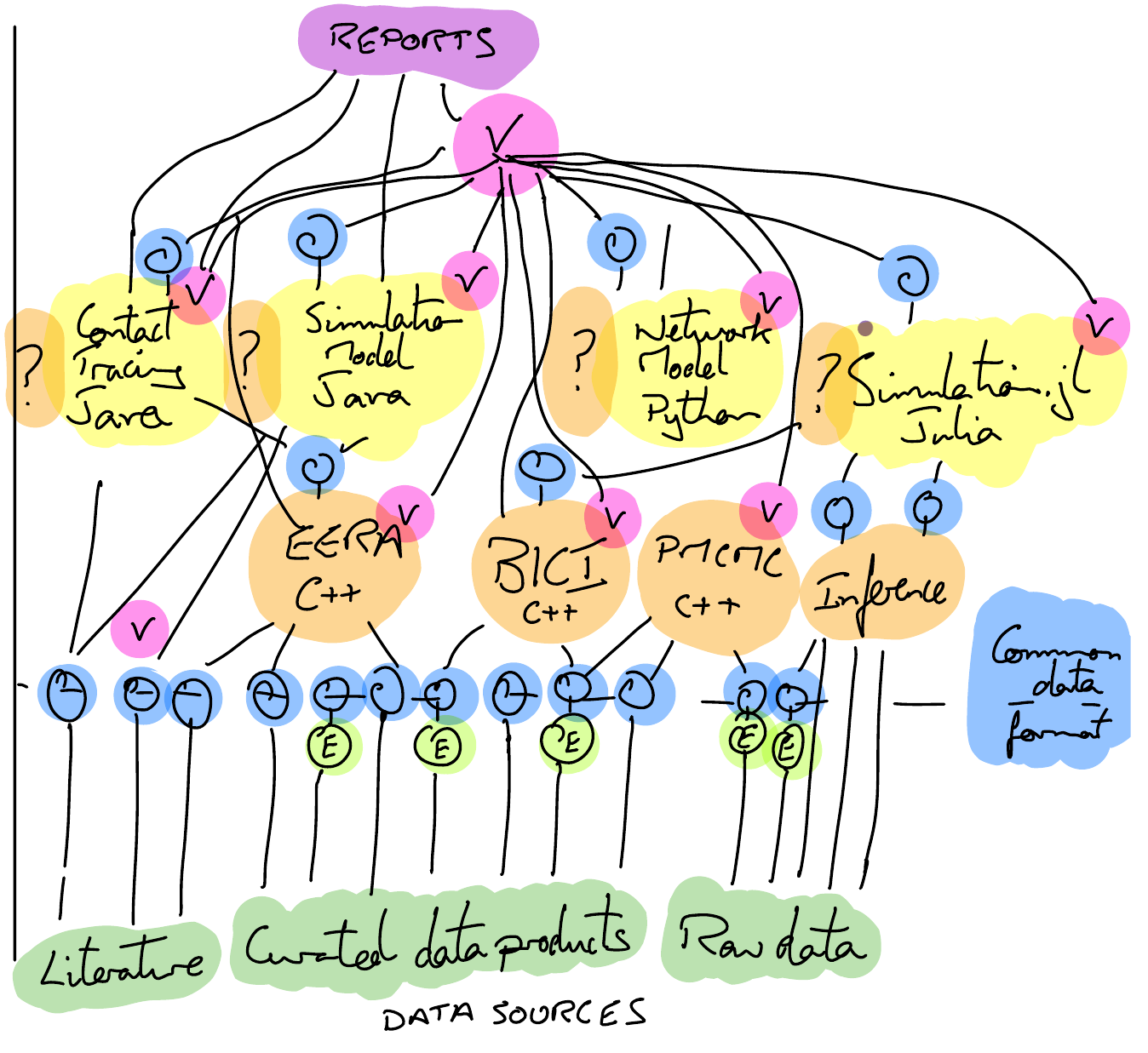}\\
        \small{(a) before discussion} & \small{(b) after discussion}
    \end{tabular}
    \caption{Two sketches illustrate the major change of the role of VIS in the SCRC modelling workflow during the initial discussion. The symbols ``V'' in pink circles indicate the needs for VIS.}
    \label{fig:Sketch}
    \vspace{-4mm}
\end{figure}

On 14 May, Dr. Richard Reeve, the SCRC modelling coordinator, first met a VIS scientist. They discussed the SCRC's overall requirements for visualisation. As shown in the first sketch in Figure \ref{fig:Sketch}(a), SCRC initially only required assistance for visualising the results of modelling, reflecting the widespread perception of visualisation as a tool only for information or knowledge dissemination. The VIS scientist described how different VIS techniques could enable domain experts to observe data quickly, analyse data with the aid of data mining algorithms, and improve their models through, e.g., visualisation of ensemble data, parameter space, and results from sensitivity or uncertainty analysis. Dr. Reeve embraced the idea of integrating VIS techniques throughout the modelling workflow, and revised the original sketch soon after (Figure \ref{fig:Sketch}(b)). The VIS scientist (referred to as VIS coordinator hereinafter) indicated the need to enlist the help from many VIS experts.

The following day (15 May), the VIS coordinator sent an email call for VIS volunteers to many VIS scientists, researchers, and developers in the UK, some of whom forwarded the call to others. By June 1, 22 VIS volunteers (including the VIS coordinator) answered the call. There are 19 faculty members, two industrial researchers, and one academic research officer. Among them, 14 indicated being able to prototype VIS software, and seven indicated willingness to engineer VIS systems.  By June 2, the coordinator held meetings with all 21 volunteers individually or in small groups.
The VIS volunteers had been using a diverse range of programming platforms. The most common denominator is D3.js \cite{Bostock:2011:TVCG}. Five VIS volunteers had experience of coding in D3.js (one became unavailable a few weeks later).

At that time, several teams in SCRC were working on six different epidemiological models and one team on inference and model validation, while substantial effort was devoted to the development of a data infrastructure for storing modelling results as well as captured data related to COVID-19 spread in Scotland. For the VIS volunteers, there were many unknowns, such as what data might be available, how it may be retrieved, what were the requirements of individual domain experts and individual models, and so on.

While the agile methodology in software engineering \cite{Larman:2003:book} and the nested model \cite{10Munzner} in visualisation advocate the necessity of iterative requirements analysis and software evaluation, they do not prescribe a full requirements analysis and software evaluation within a single iteration. Otherwise, they would be similar to the waterfall methodology. In the VIS literature, many application papers indicate that it usually takes many months to acquire a meaningful set of requirements (e.g., six months in \cite{AbdulRahman:2013:CGF,Fang:2017:TVCG} and 12 or more months in \cite{Lloyd:2011:TVCG,Elshehaly:2021:TVCG}).
To support an emergency response, a lengthy delay due to requirements analysis would not be acceptable. Hence we had to complement user-centred requirements analysis with the existing knowledge documented in the VIS literature, and commenced the development as soon as we had understood a partial set of requirements.

As reviewed in Section \ref{sec:RelatedWork}, many papers in the literature reported VIS techniques and tools for supporting healthcare applications, model development, and mission-critical operations. If one can identify the data types, user tasks, and user knowledge in an application, one can relate them to the requirements in previously reported applications that featured similar characteristics of data, tasks, and users. During the two weeks when we were recruiting VIS volunteers, we gained our understanding of:
\begin{itemize}
    \item \emph{Datatypes} --- Based on several briefs from the SCRC modelling teams, we quickly learned that there would be a huge amount of time series data, and some geographical data (e.g., maps), network data (e.g., contact tracing), and multivariate data (e.g., demographic data). Building on our knowledge of VIS literature, we anticipated that some other types of data that might result from analytical algorithms, such as similarity matrices.%
    \item \emph{User tasks} --- Building on our knowledge of other VIS applications, we quickly established that there would be a need for viewing time series in different ways for observation and comparison, in order to evaluate a model run against captured data, other runs, and other models. We anticipated that some analytical tasks would benefit from data mining algorithms, and at a later stage, domain experts would become interested in ensemble data visualisation and parameter optimisation.%
    \item \emph{User knowledge} --- Building on our experience working with other domain experts, we anticipated that (i) domain experts were highly knowledgeable about their own models, but could not avoid frequent observation of captured data and model results; and (ii) they were familiar with the major geographical locations in Scotland, but would need to incorporate map-based visualisation for smaller regions in Scotland and other UK regions.
\end{itemize}

Meanwhile, we also consulted the abstracted theories and methodologies in the VIS literature. We observed that there was a need for all four levels of visualisation \cite{Chen:2016:TVCG}. As illustrated in Figure \ref{fig:Sketch}, although \emph{disseminative visualisation} was the initial requirement, the above considerations about data, tasks and users confirmed that the priority should be given to \emph{observational}, \emph{analytical}, and \emph{model-developmental visualisation}. 

The VIS volunteers were thus organised into several teams, including a generic supporting team (focusing on observational visualisation), an analytical support team, four modelling support teams, and a disseminative visualisation team. We placed all D3.js developers into the generic support team, and distributed other VIS volunteers according to their expertise and time capacity. In the following four sections, we report the activities of these teams, including further requirement analysis conducted by each team.%

\begin{figure*}[t]
    \centering
    \includegraphics[width=176mm]{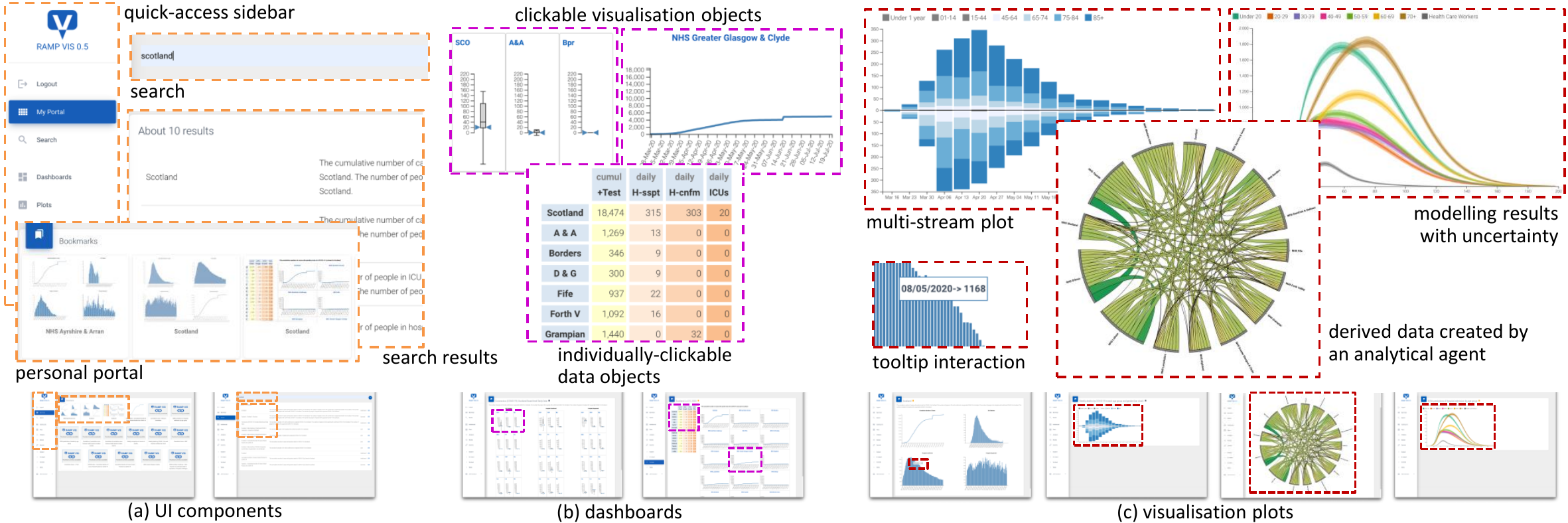}
    \caption[]
    {The RAMP VIS server (vis.scrc.uk) provides users with a user interface (with search facility, personal portal, etc.), a collection of dashboards (with clickable visual and data objects), and various plots for visualising the data hosted by the SCRC data infrastructure.}
    \label{fig:UI}
    \vspace{-5mm}
\end{figure*}

\section{Generic Support and RAMP VIS Infrastructure}
\label{sec:Generic}
\textbf{Infrastructure Setup.} \quad
The \emph{generic support team} consists of mainly VIS volunteers who can program in D3.js and have developed and deployed VIS systems.
Our requirement analysis indicated that domain experts were not able to observe data regularly. Using the recently-proposed method for optimising VIS workflows \cite{Chen:2019:CGF}, we quickly identified that this was caused by the cost of reading and visualising data using spreadsheets, and a solution is to develop a VIS infrastructure closely coupled with the SCRC data infrastructure that was being developed. The goal of the team was to enable observational visualisation for every piece of data held by the data infrastructure. 
  
The development of the SCRC data infrastructure started several weeks before VIS volunteers joined SCRC. A group of professional research software engineers (also volunteers) have carried out the design and implementation since. The goal is to capture the provenance of models and their results, enabling all contributing elements traceable from results to models and the conclusions drawn.
Transparency is thus a key principle.
All models and core software components are open-source \cite{SCRC:2020:git}.
Hence the VIS infrastructure has also been developed in the open.

The UK Science and Technology Facilities Council (STFC) provides the data and VIS infrastructure featuring virtual machines on the STFC cloud service, including a chat platform for collaboration and a data registry for web applications. The readiness of STFC for emergency responses enabled the hardware for the VIS infrastructure to be available within 24 hours after our request.

As emergency responses, the SCRC data infrastructure, VIS infrastructure, and six epidemiological models were developed in parallel. While the generic support team was waiting for the measured data, we had access to some Scotland data in three spreadsheets, which contained over 300 time series and a few data tables. We anticipated that there would be at least thousands of time series when data from other regions and model runs became available. Such scale would be a challenge to the domain experts as well as the VIS developers. Users would need to assess the relevant plots quickly, while developers would need to adapt each visualisation program (referred to as a \emph{VIS function}) to other applicable data with minimal development effort.

\noindent\textbf{User Interface.} \quad
To address the need of the domain expert users, the RAMP VIS server provides the following facilities (Figure \ref{fig:UI}):
\begin{itemize}
    \item A user interface (UI) with a side bar for accessing visualisations organised in categories;
    \item A multi-keyword search facility;
    \item A personal portal for storing frequently-used visualisations;
    \item Because each visualisation is given a unique URL, users can also tag frequently-used visualisations on a web browser;
    \item A collection of dashboards, each providing links to other dashboards and visualisation plots. 
\end{itemize}

On the RAMP VIS server \cite{RAMPVIS-UI:2020:git}, there are broadly two types of visualisations, \emph{dashboards} and \emph{plots}. Each dashboard is designed to show key indicators and/or summary plots that some domain experts need to view frequently. For example, it may show the daily data of a region or a summary of a model run (Figure \ref{fig:UI}(b)). Some information may serve as overviews while others serve as detailed views.
The data objects, visual widgets, or summary plots on a dashboard are all clickable, providing a gateway to fuller or more detailed visualisations or other dashboards when required. In addition to a set of pre-defined dashboards, the generic support team provides a service to SCRC domain experts for constructing new dashboard whenever needed.

The team have developed a variety of visualisation plots. As shown in Figure \ref{fig:UI}(c), some plots feature interactive capabilities, some compare multiple data streams, some convey the analytical results produced by data mining algorithms, and some display modelling results with estimated uncertainty.  

\begin{figure*}[t]
    \centering
    \includegraphics[width=172mm]{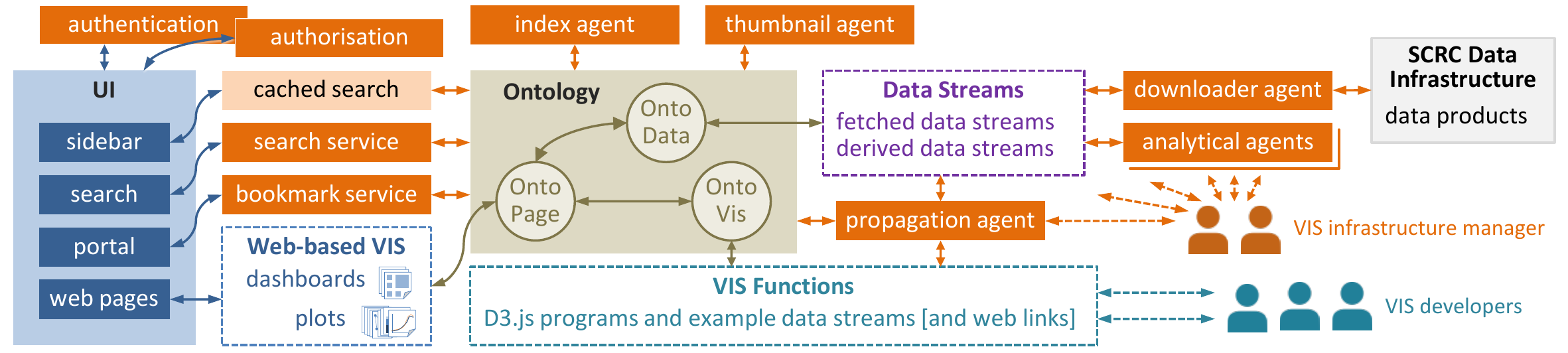}
    \caption{The architecture of the RAMP VIS server, featuring its ontology, agents, and services, and their relationships with the UI, web pages (dashboards and plots), VIS functions (programs and test data), fetched and derived data streams, and SCRC data infrastructure.}
    \label{fig:Architecture}
    \vspace{-4mm}
\end{figure*}

\noindent\textbf{Architecture and Ontology.} \quad
To address the aforementioned second need for propagating each VIS function (either a dashboard or a plot) to other applicable data, the team has designed and developed an ontology- and agent-based architecture for the RAMP VIS server \cite{RAMPVIS-API:2020:git}.
When a VIS developer fetches a task of writing a VIS function in D3.js, an infrastructure manager creates a program template with one or more example data streams.
The binding of the VIS function and the given data streams results in a unique web page.  
Once the development completes, the VIS function can be reused by replacing the sample data stream with other applicable data streams.
Through a simple UI, the infrastructure manager queries a \textbf{Search Service} to find all suitable candidate data streams, finalises a collection of data streams for propagation, and calls a \textbf{Propagation Agent} to create new bindings (and web pages) for these data streams automatically.

As illustrated in Figure \ref{fig:Architecture}, in the VIS infrastructure, an ontology provides the vital support to the search facilities that enable users to find desired dashboards and plots and the infrastructure manager to find applicable data streams for propagation.
The ontology is a graph data structure that stores the relationships among all VIS functions, all data streams, and all data-VIS bindings (and the resulting web pages).
Because we modelled the ontology using a document data model \cite{chodorow2013mongodb}, we implemented the ontology using three MongoDB database collections, which are:
\begin{itemize}
    \item \textbf{OntoVis} for defining and keeping the records of all VIS functions and their metadata;
    \item \textbf{OntoData} for storing the records of all data streams and their metadata;
    \item \textbf{OntoPage} for maintaining the binding points between VIS functions and
    data streams, there metadata and the URLs of the corresponding web pages.
\end{itemize}

As shown in Figure \ref{fig:Architecture}, in addition to the aforementioned \textbf{Search Service} and \textbf{Propagation Agent}, there are:
\begin{itemize}
    \item A \textbf{Downloader Agent} for fetching data from dynamically-changing data automatically from the SCRC data infrastructure;
    \item A set of \textbf{Analytical Agents}, each applies an analytical function or data mining algorithm to a predefined collection of data streams and generates derived data to be rendered, e.g., a similarity matrix of a collection of time series (see also Section \ref{sec:Analytical});
    \item An \textbf{Indexing Agent} that periodically scans the database operation logs and updates various textual descriptions in the ontology, which may be used in search or displayed by VIS functions;
    \item A \textbf{Thumbnail Agent} for creating and updating the thumbnails of dashboards and plots that may change due to the dynamic change of the underlying data.
    \item A \textbf{Bookmark Service} for managing bookmarks in users' portals;
    \item An \textbf{Authentication Service} for approving a user's login action;
    \item An \textbf{Authorization Service} for distinguishing ordinary or administrative users.
\end{itemize}

The RAMP VIS server was implemented with backend microservices using two state-of-the-art REST-API frameworks: JavaScript-based NodeJS and Python-based Flask. The I/O-intensive operations (e.g., database or file-system access) are performed asynchronously and are implemented with a NodeJS microservice. The CPU intensive operations (e.g., running \textbf{Analytics Agents}) are implemented with a Flask microservice. The Flask framework also provides \textbf{Analytics Agents} with some off-the-shelf analytical libraries, e.g., NumPy, SciPy, scikit-learn.
The SCRC data infrastructure provides the \textbf{Downloader Agent} with Python APIs to fetch H5 data.

\begin{figure*}[t]
    \centering
    \includegraphics[width=0.95\textwidth]{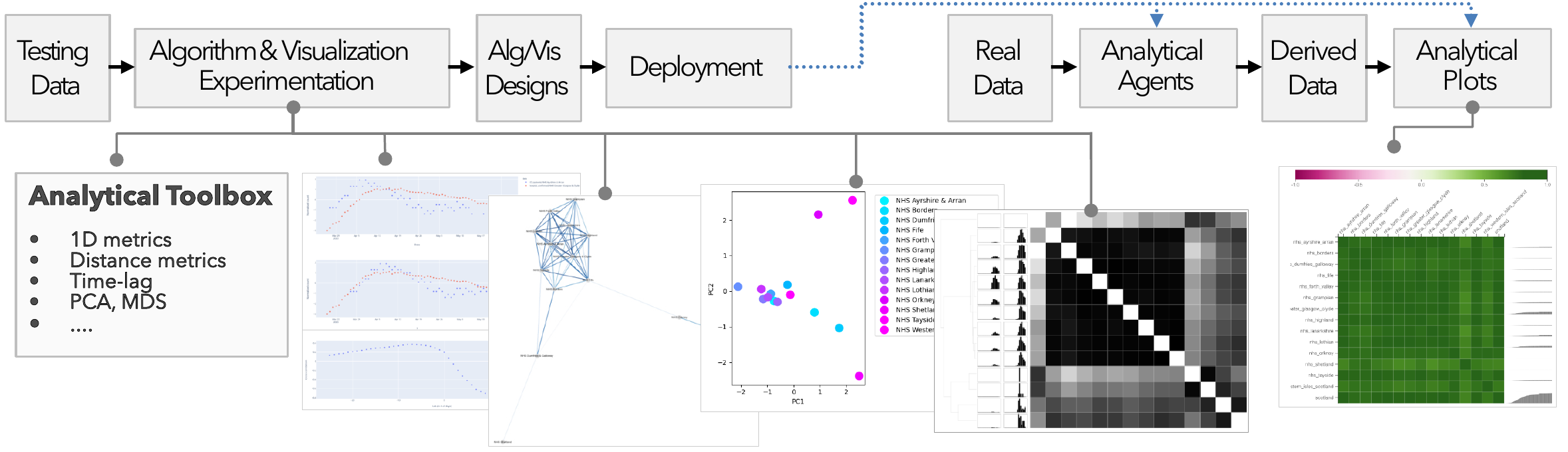}
    \caption{The analytical support team develops and experiments with different analytical algorithms and visual representations (top left) and selects some designs to be deployed on the VIS infrastructure by the generic support team (top right).
    The lower part of the figure shows examples of analytical visualisations for time series analysis, including time-lag plot, force-directed graph, scatter plot, and heatmap matrix.}
    \label{fig:Analytics}
    \vspace{-5mm}
\end{figure*}

\section{Analytical Support}
\label{sec:Analytical}
As exemplified by numerous visual analytics papers,
VIS applications with a large data repository are expected to employ both data analysis algorithms and visualisation techniques.
For example, analytical tasks have been critical for assessing model performance and uncertainty (e.g., comparing predictions to observations and comparing multiple model runs)~\cite{konyha2006interactive}, and for exploring epidemiological data (e.g., identifying areas, time periods, or demographic groups exhibiting similar trends in outbreak progression)~\cite{bhaskaran2013time}.
With the huge number of time series to be hosted by SCRC data infrastructure, we anticipated that simply relying on observational visualisation might not be efficient or effective. We thus grouped several VIS volunteers with strong data mining experience into the \emph{analytical support team}.

After several attempts to acquire detailed requirements for analytical visualisation capacity, we learned a high-level requirement from some modelling scientists, i.e., ``cross-model validation'' would be needed at some stage. With their expertise and initiatives, the team members anticipated that comparing time series would be an unavoidable analytical need because of the sheer volume of dynamically expanding time-series data.
The initial data available to the team contained hundreds of time series for different regions of Scotland, different indicators (e.g., test, case, hospitalised, and fatality), different genders and age groups, and so on.
While waiting for modelling data to be prepared for comparative analysis, the team started to develop visual analytics techniques for summarising, simplifying, and comparing time series and for searching and visualising patterns and structures in the data.
Building on the VIS literature on time-series~\cite{aigner2011visualization},
we anticipated the following analytical tasks: (i) discovering recurring trends, (ii) looking for outliers, (iii) identifying clusters, and (iv) measuring similarities according to one or more characteristics (e.g., scale, gradient, time lag, etc.).
The team's anticipation was confirmed when more data and specific requirements arrived several months later.

Guided primarily by these tasks, we experimented with a number of \textit{analytical visualisations} in conjunction with an \textit{analytical toolbox} that consists of different metrics and algorithms for quantifying the characteristics of individual time series, computing pairwise similarities, and transforming the time series to feature spaces that enable their similarities and clustering to be visualized.

\noindent\textbf{Analytical Toolbox.} \quad
We started developing the analytical toolbox by creating a library of low-level analytical filters that treat time series as 1-D signals. Since time series may contain noise, various types of filters (e.g., flat, Hanning, Hamming, Bartlett, and Blackman windows) can be used to smooth time series when required. We then added a comprehensive library of analytical metrics for measuring the distance, difference, similarity, or error between two time series (e.g. mean square error and many of its variations, Pearson correlation coefficient, structural similarity index measure, mutual entropy, Spearman correlation coefficient, Kendall's tau, peak signal-to-noise ratio, $F$-test, and so on).
We further included algorithms such as dynamic time warping (DTW) and dimensionality reduction methods such as principal component analysis (PCA) and multi-dimensional scaling (MDS).

\noindent\textbf{Analytical Visualisations.} \quad
We also experimented with a number of visual representations, focusing on comparing $N$ time series with $T$ data points.
If one needs to determine any group of $k$ segments of time series that may be similar, the number of possible groups to be observed would be at the level of $O(N^kT^k)$,
hence using analytical algorithms to narrow down the search space is highly beneficial \cite{Chen:2016:TVCG}.
However, relying on metrics alone is not sufficient since time series could be similar/dissimilar due to factors not encoded in the data (e.g., differences in terms of demography and intervention). In that respect visualisation provides ways for domain experts to incorporate their knowledge when analysing and comparing different time-series.
Figure~\ref{fig:Analytics} shows one set of our experiments for analysing the time series associated with the 14 regional health boards in Scotland.
\begin{itemize}
    \item A \textit{time-lag visualisation} two compares time series by registering them using a cross-correlation that computes the displacement of one time series relative to the other. A viewer can foresee what the future may look like in one board if it follows the same trends as another board, but with a delay.
    \item A \textit{heatmap matrix} shows pairwise similarity scores among all $N$ time series reported by different health boards. Row or column headers can be accompanied by time-series profiles for detailed observations. This is especially important when the similarity/difference measures are difficult for viewers to interpret.
    \item A \textit{force-directed graph} produces a layout where the $N$ nodes represent $N$ time series, and the length of each edge encodes the similarity/difference (short/long) between a pair of nodes.
    This visual representation is particularly useful for users to discover clusters of similar time series and outliers.
    \item A \text{chord diagram}, which is shown in Figure \ref{fig:UI}, places $N$ time series as segments/nodes along a circle, and uses the thickness or color of each chord to encode the similarity/difference measures. 
    \item A \emph{scatter plot} compares $N$ time series in their feature space. Typically, two most important features (e.g., principal components computed using PCA) are selected as the axes of a 2-D space, and each time series is positioned as glyphs in the space according its feature coordinates.
\end{itemize}

When the analytical support team examined these experiment results with the domain experts, one domain expert commented ``These give us a lot to think about. It is not that we do not require these. We just overwhelmed by what visualisation can do.''

\noindent\textbf{Infrastructural Support.} \quad
As shown on the right of Figure \ref{fig:Analytics}, following the experiments, the analytical support team selects analytical algorithms and visual representations to be integrated into the VIS infrastructure maintained by the generic support team. Each analytical algorithm becomes an \textbf{analytical agent}, while each visual representation becomes a plot.
Because many data streams in the infrastructure are updated dynamically, each \textbf{analytical agent} is scheduled to recompute various measures and generate derived data automatically. In this way, when an analytical plot is called, it always displays the analytical results based on the latest data.

\section{Modelling Support}
\label{sec:Modelling}
Mathematically, finding an optimal model to forecast the contagion patterns of COVID-19 in
different conditions (e.g., geographical, social, seasonal variations, different human interventions, etc.) is an intractable problem. Nevertheless the effort to develop better models and improve existing ones is both necessary and desirable. 
When VIS volunteers were first gathered together, we anticipated that supporting the model development in SCRC would be the most challenging undertaking, because VIS would have to support the search for better models in an NP space \cite{Chen:2016:TVCG}. We therefore organised some 10 VIS volunteers into four \emph{modelling support teams} (referred to as Teams $\mathbf{M}_A$, $\mathbf{M}_B$, $\mathbf{M}_C$, and $\mathbf{M}_D$ below), providing opportunities for each team to focus on supporting one or two SCRC modelling teams through close collaboration.

\subsection{Team M$_A$: Supporting 1-km$^2$ Spatial Simulation}
\label{sec:ModellingA}

\begin{figure}[t]
    \centering
    \includegraphics[height=54mm,trim=60px 80px 320px 150px, clip=true]{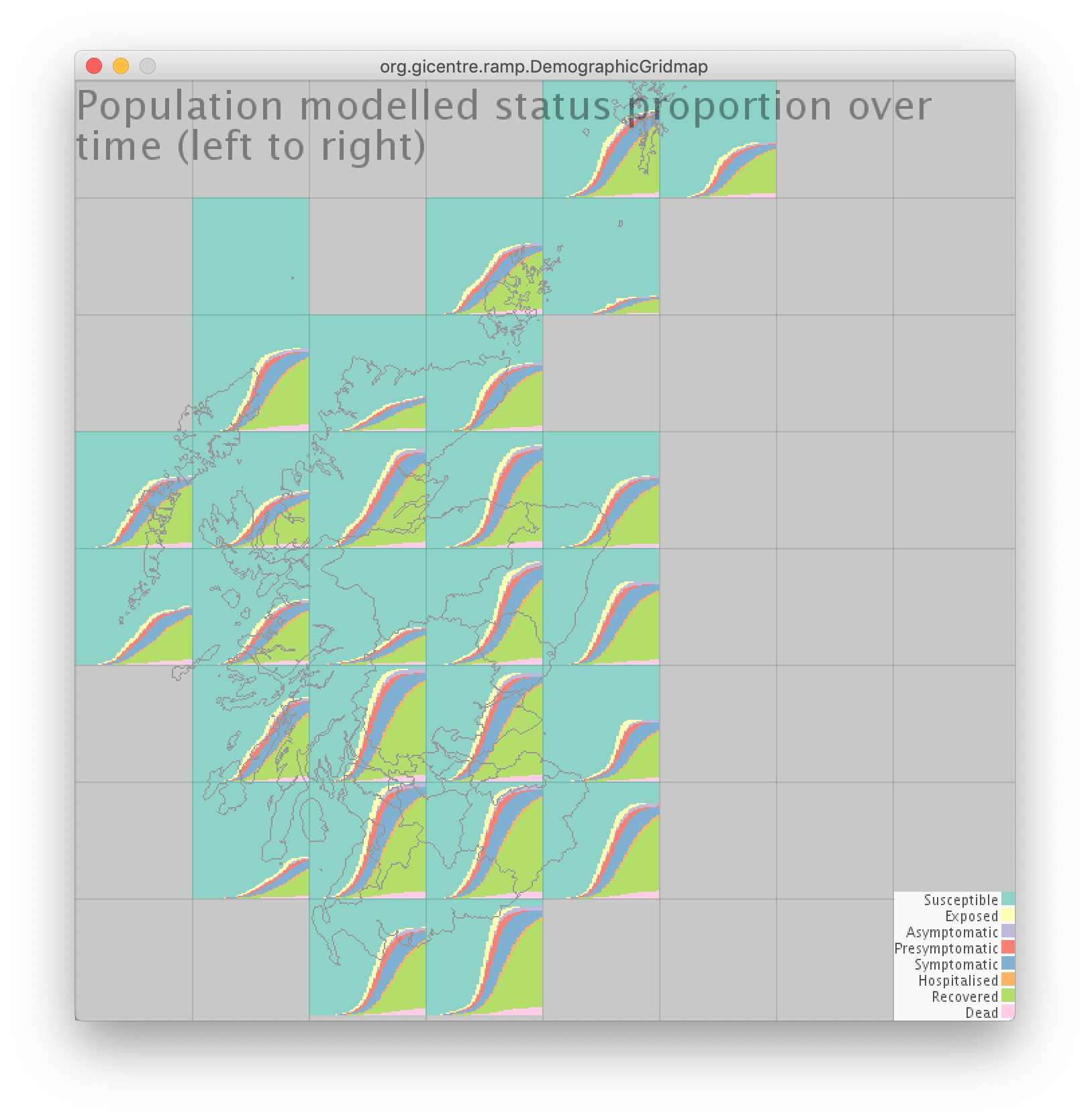}
    \includegraphics[height=54mm,trim=60px 80px 250px 150px, clip=true]{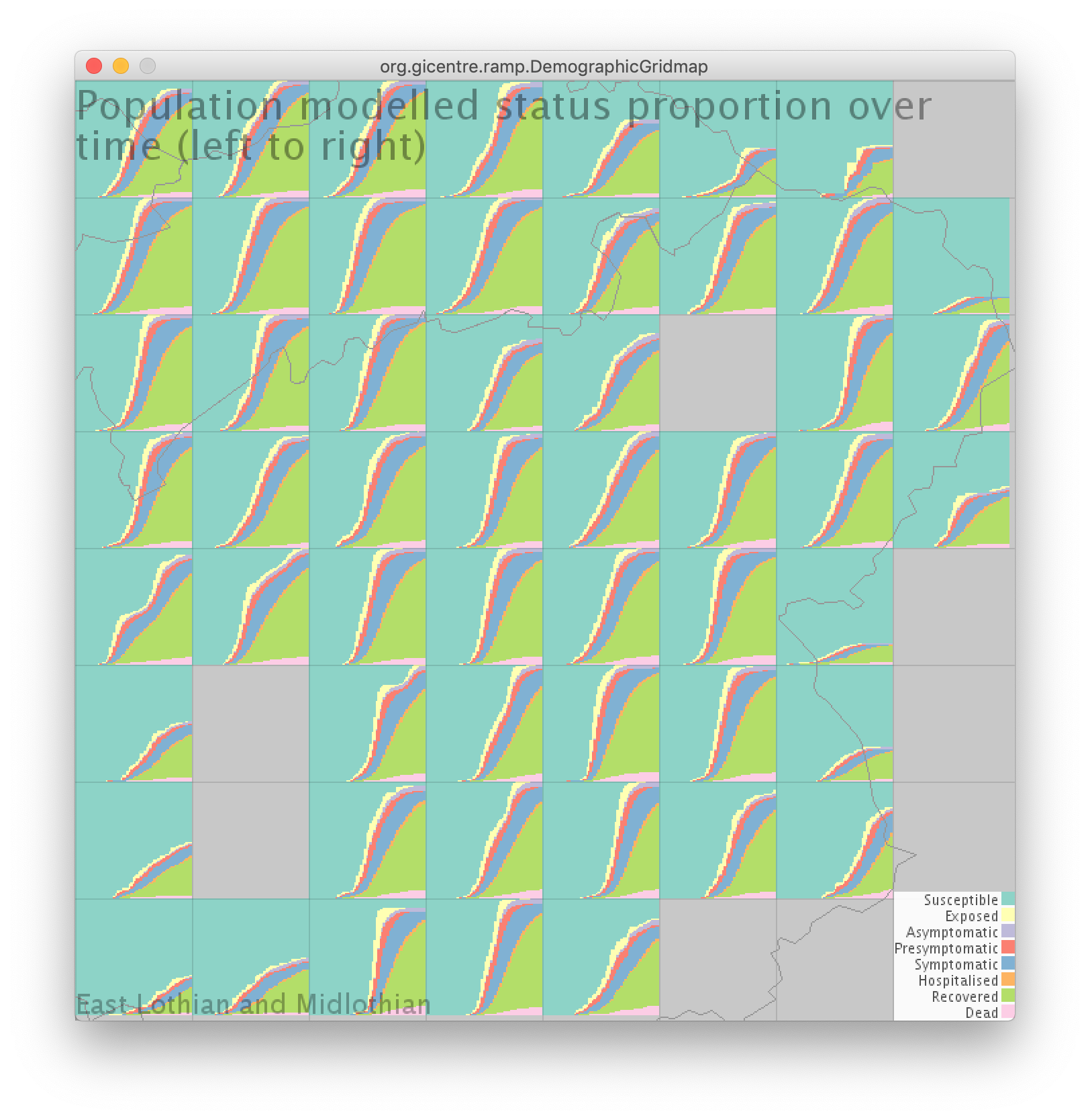}
    
    \includegraphics[height=54mm,trim=60px 80px 320px 150px, clip=true]{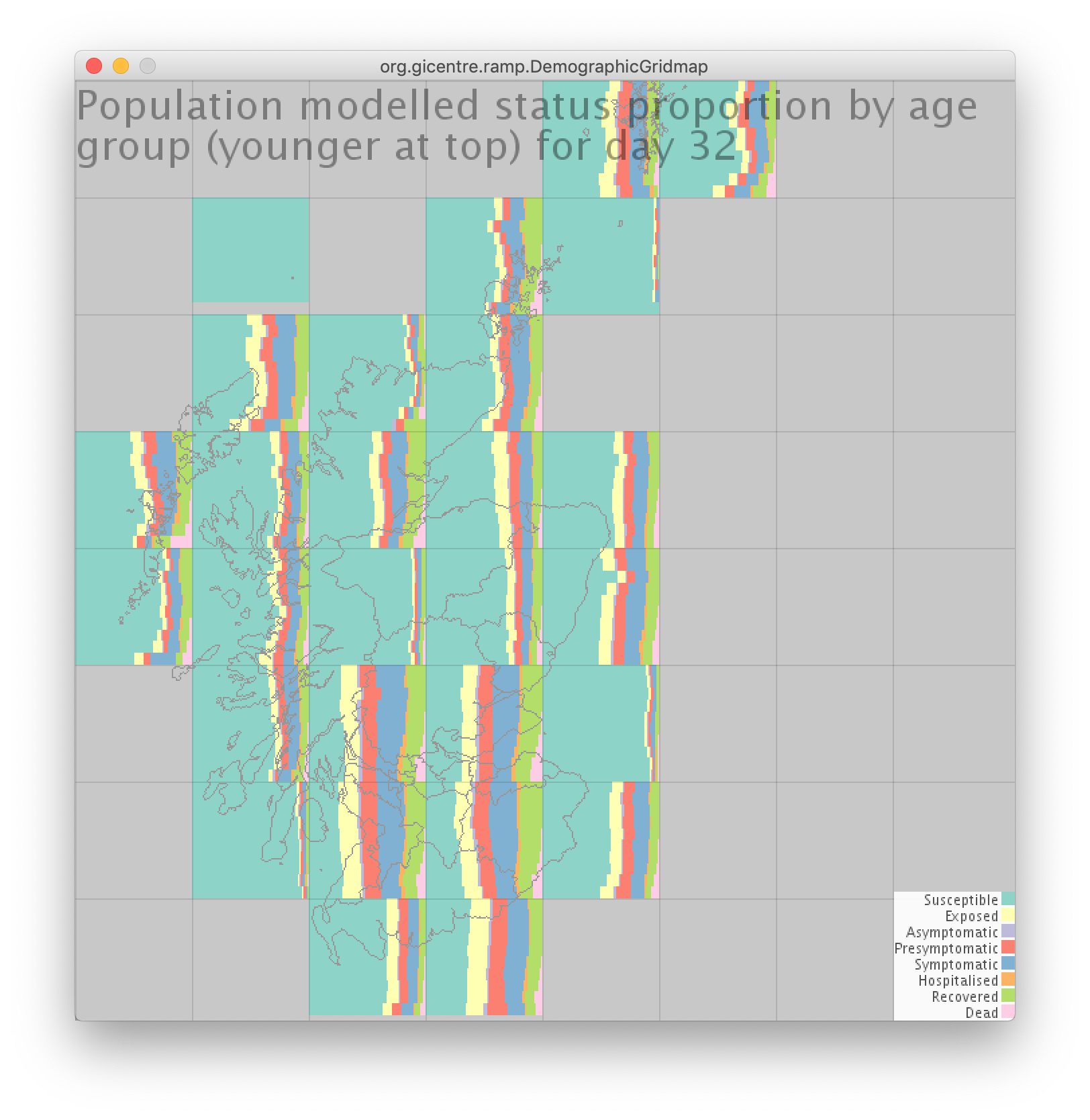}
    \includegraphics[height=54mm,trim=60px 80px 250px 150px, clip=true]{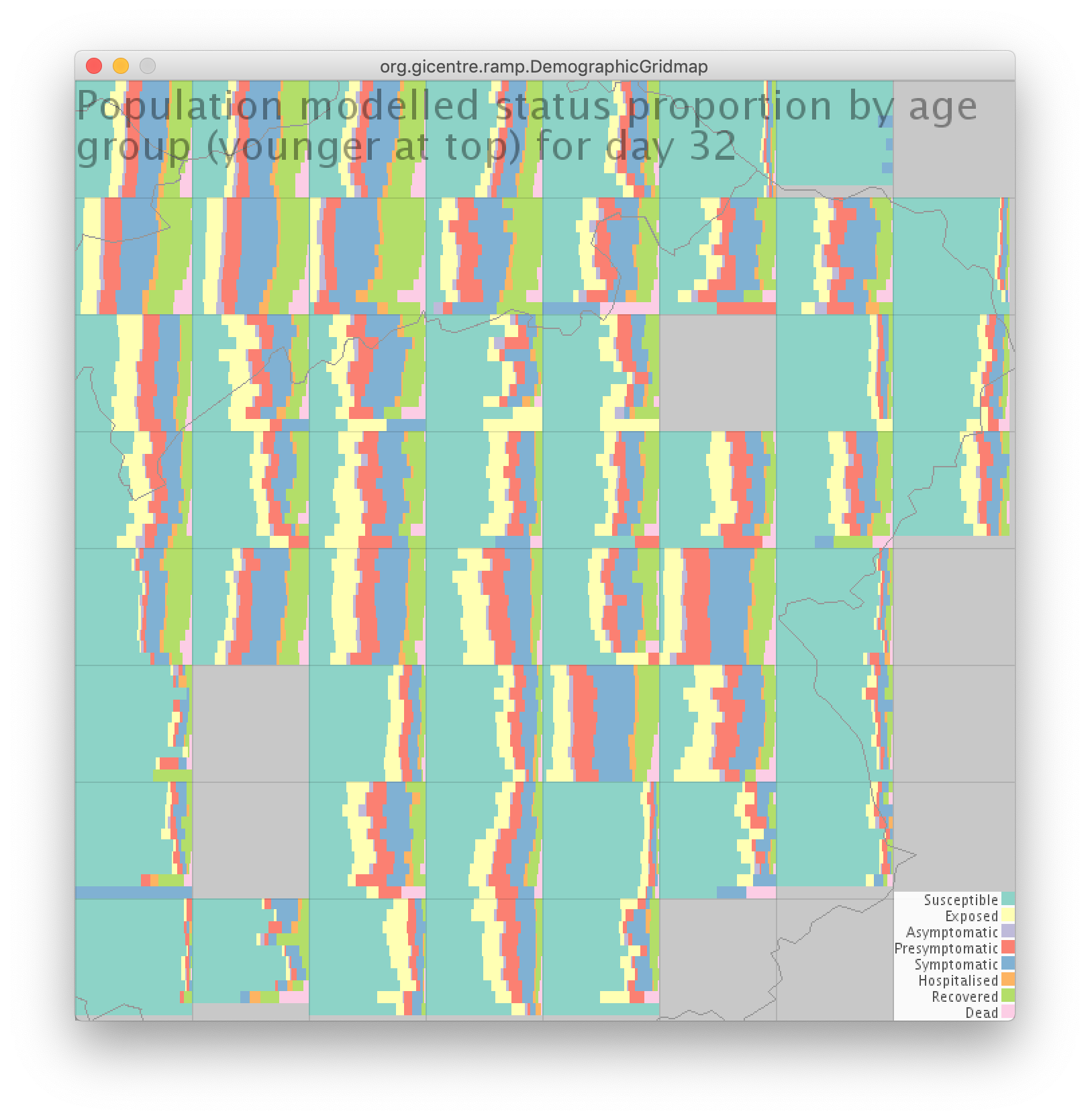}
        
    \vspace{2mm}
    
    \includegraphics[width=85mm,trim=0px 0px 0px 0px, clip=true]{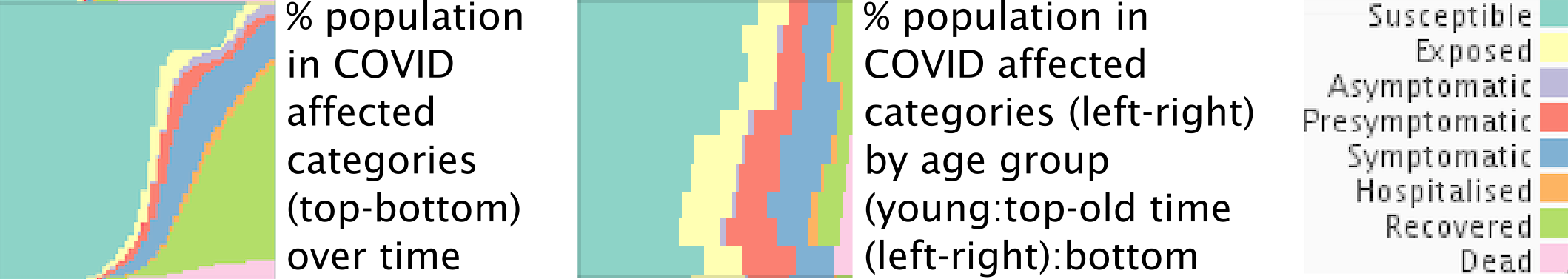}    
    \caption{Model output showing proportion of population by COVID-affected category for the whole of Scotland (left) and more detail in SE Scotland (right). Top: glyphs show change in proportion of COVID-affected population (top-bottom) from day 0 (left) to day 60 (right). Bottom: proportion of COVID-affected population (left to right) for different age groups (young-old; top-bottom).}
    \label{fig:ModellingA}
    \vspace{-5mm}
\end{figure}

Model \emph{Simulation.jl} \cite{Simulation.jl:2020:git}
simulates the spread of COVID-19 based on spatial proximity and its effect on the local population according to its demographic structure, over time. The inputs are population counts in 10-year age-bands in 1$\times$1km$^2$ to 10$\times$10km$^2$ grid cells across Scotland. Given an initial set of ``seed'' locations on day 0, the model outputs the number of people in different COVID-affected categories for each subsequent day by age group in the same grid cell. Figure~\ref{fig:ModellingA} shows a set of simulation results.

We met the domain experts as the first version of the model was being created. There was an urgent need to visually inspect the relative proportions of COVID-affected individuals in different categories over time and space. When discussing the high-resolution model outputs and strategies one might use to summarise them to validate model outputs and (later) to compare different modelling scenarios, the need to freely explore these prior to establishing fixed tabular summaries became apparent.
As this need was so urgent, we quickly established two VIS requirements: to enable (i)~studying the relative proportions of COVID-affected individuals in different categories over space and time, and (ii)~exploring the results at different scales from Scotland-wide to 1-km$^2$ neighbourhoods.

Our solution to these challenges was to use interactive ``tilemaps'' \cite{city20884} (also known as ``glyphmaps'' \cite{https://doi.org/10.1002/env.2152} and ``embedded plots'' \cite{doi:10.1080/10618600.2014.896808}) with (a)~glyphs representing multiple aspects of the modelled output together and (b)~on-the-fly interactive gridding of the output at a suitable resolution in response to zoom/pan user interaction. The technical challenge was to make interaction, with this very high-resolution data quick and responsive enough to facilitate, rather than impede, exploration. 

On the left of Figure~\ref{fig:ModellingA}, data at a coarse spatial resolution is superimposed on a map of Scotland. The glyphs in the top, left image show the aggregated temporal trend as the population moves through the COVID-affected categories, with significantly lower proportions of affected populations in the lower populated areas of NW Scotland. The image below this is a snapshot on day 32, showing the different rates at which the virus is affecting the population. The right column is a more detailed view of SE Scotland, as a result of zooming/panning. The snapshot on day 32 (bottom right) shows that the virus is affecting different age groups differently. This is largely due to differences in resident population structure. Domain experts with knowledge of population density can see that low-density areas seem to act as ``firebreaks''. Although this is the inner working of the spatial spreading algorithm, its appearance in the visualisation started a debate among the domain experts, influencing the next stage of model development -- to investigate the importance of population density on speed of disease spread.
The interactive tilemaps with glyphs have provided a basis for the ongoing work for evaluating the relative importance of different factors in the modelling and comparison of different lockdown scenarios.

\vspace{-2mm}
\subsection{Team M$_B$: Supporting Simple Network Simulation}
\label{sec:SNS}


\begin{figure}[t]
    \centering
    \includegraphics[width=65mm]{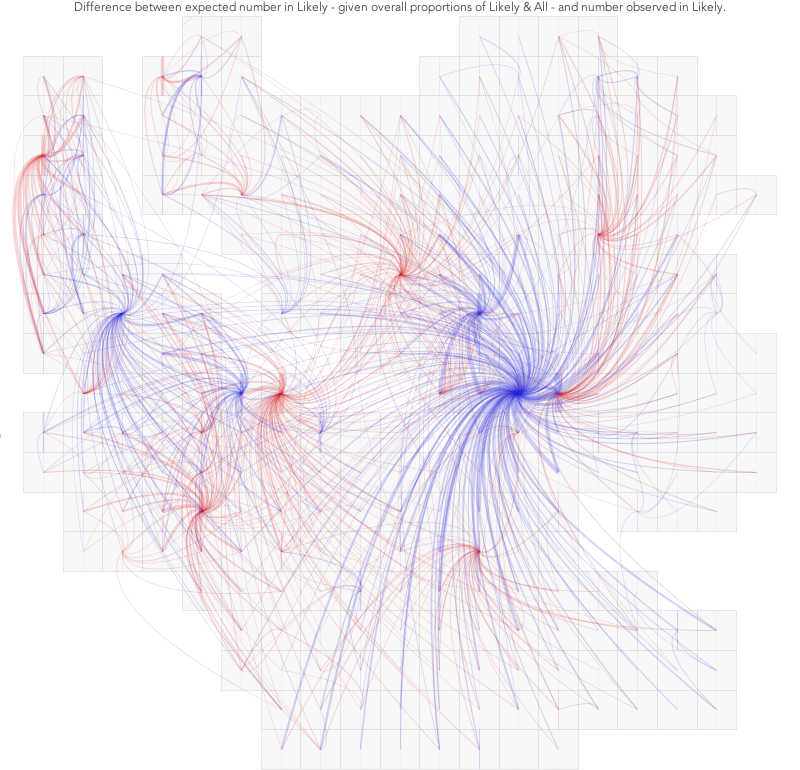}
    \caption{Largest daily flows between MSOAs in Glasgow on a transformed GridMap.
    Certain destinations clearly have flow signatures that are greater (red) or less (blue) than the proportions predicted by regular daily flows under this `lockdown' scenario.}
    \label{fig:ModellingB:output}
    \vspace{-5mm}
\end{figure}

The Simple Network Simulation (SNS) is a disease state progression model with multiple potential inputs and outputs \cite{SNS:2020:git}. The former include spatial models of social interaction at a variety of scales, the latter varying according to model parameters and by time, age-group and geography.
When we connected with the SNS team, they were considering simulating commuting patterns during lockdown by grouping daily flows for those in employment sectors differently effected by the ``work from home'' edict of April 2020.
Their graphical approach used origin-destination matrices of data recorded in the UK Census of Population.
However, flow quantities, group differences, and any geographic variations or effects of scale were difficult to see (Figure~\ref{fig:ModellingB:output}).

We began with a design challenge and used this to immerse ourselves in the problem \cite{hall2019design} -- learning about the SNS model, gaining trust and finding other problems to which effective visualisation might be usefully applied.
Our frequent, rapid feedback mechanisms encouraged creative thinking \cite{goodwin2013creative,kerzner2019framework} and discourse around data \cite{beecham2020design}. Involving video-calls, emoji-filled chat streams and structured design documents, we might characterise them as mini BIE loops \cite{mccurdy2016action} in which \emph{redesign} might result in a shift in emphasis ranging from zero to a total switch of channel~\cite{wood2014moving}.
This approached \emph{lockdown immersion} \cite{hall2019design},
through which we 
\emph{enriched} our understanding of the modellers domain,
\emph{explored} new spaces for design and enquiry
and \emph{built} relationships within and beyond the SNS team.

This enabled us to collectively identify and visually explore two specific domain questions: (a) which types of workers should we include in the input network (and what difference does this make to outputs)? (b) what do model outputs look like (and how do they vary over time and by age-group)?
Further questions that we hoped to address were:
(c) how do the answers vary with scale and geography? (d) is visualisation effective in answering these questions?

The discourse resulted in some preliminary answers to these questions.
For instance, for questions (a, c), the nature of the input network has an effect on outputs, with distinct spatial variation and greater effect at smaller scales.
We redesigned graphics, allowing us to identify the areas most affected by the industry factor, areas with high out-of-area epidemiological importance, and the sources of those who visit them (Figure~\ref{fig:ModellingB:output}). But they do not dictate the scale or content of the network we should use. This resulted in a better informed input network and an emphasis on smaller scale ``data zones''.
For (b, c), model outputs were not particularly varied by space, time or attribute, but we do have plausible candidate methods that will allow such variation to be detected and assessed when the SNS models are finalized.
Importantly, the engagement also enables us to be more confident about injecting visualisation into the modelling process, to shine light on the models as they are developed, tested, and parameterised, in ways that had not been considered by the modellers (Figure~\ref{fig:Sketch}), and perhaps at fine scale.

\begin{figure}[t]
    \centering
    \includegraphics[width=\columnwidth]{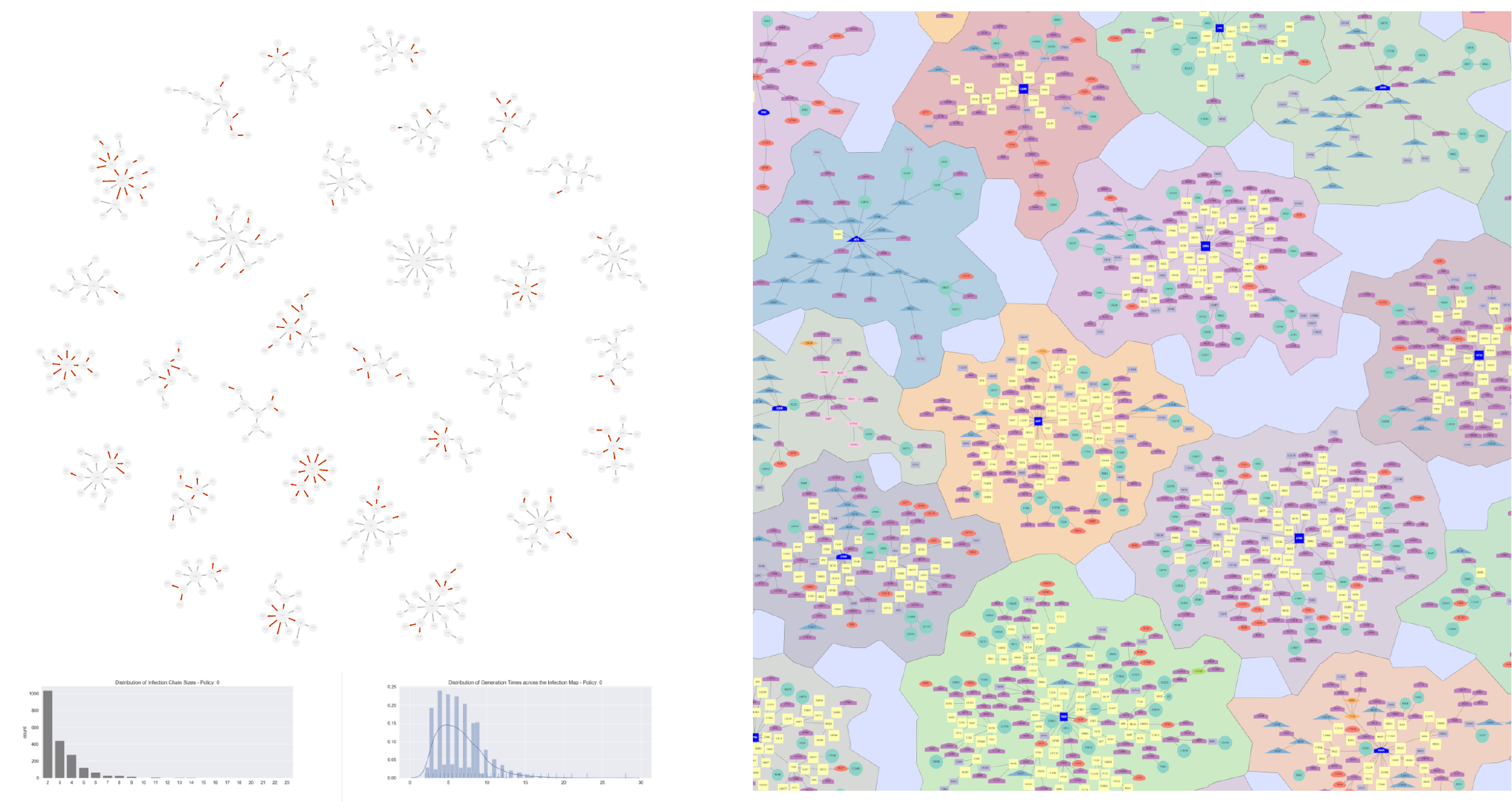}
    \caption{Some of the largest infection chains visualised -- with asymptomatic transmissions highlighted (left), and with index nodes and locations of infections expressed along with a visual indication of the sizes of the chains (right). The visual analysis was also enriched with extracted graph-theoretic metrics (left-bottom).}
    \vspace{-5mm}
    \label{fig:ModellingC}
\end{figure}

\vspace{-2mm}
\subsection{Team M$_C$: Supporting Contact Tracing Modelling}
\label{sec:contactTrace}
By the time VIS volunteers joined the SCRC, the contact tracing model~\cite{ContactTracing:2020:git} was already in development.
The model simulates the spread of COVID-19 through a dynamic network that encodes the potential contacts among millions of individuals.
These simulations result in some very large temporal networks~\cite{12Holme}.

We immediately established bi-weekly meetings with the modelling scientists.
It quickly became clear that the domain experts did not have access to bespoke network visualisation tools, and were primarily relying standard plots to view simulation results as disease progression curves and some summary metrics such as $R$. They were also comparing different intervention policies through such plots. We noticed that the temporal networks, which were used to derive the summary information, were never visualised.
Hence, the first urgent requirement for VIS was to enable domain experts to observe such networks in order to gain an intuitive understanding about the temporal and topological behaviours of the model. The visualisation would also assist domain experts in communicating modelling results and informing policy making.

We addressed the requirement by using existing network visualisation tools to minimise the delay due to software development. This allowed us to build familiarity with the model and the data, while providing example visualisations to stimulate our discussions with the domain experts.
We then progressed to more advanced VIS techniques, e.g., employing scalable graph-drawing techniques~\cite{17Simonetto,20Simonetto}, geographic-inspired metaphors~\cite{gmap}, and graph-theoretic analysis for derived metrics to complement network visualisation (Figure \ref{fig:ModellingC}).

With bi-weekly collaboration meetings providing continuous feedback and ideas, we improved our prototypes iteratively through a web-based ``project diary'' and an open software repository \cite{ContactTraceVis:2020:git}.
As the collaboration matured over the period, we observed a trend that both domain experts and VIS volunteers actively contributed to the discussions on model building and visual design together.
It became difficult to label whether a discussion was about visualisation or the model itself,
and there were more discussions on generating insight than on producing software. 

\begin{figure}[t]
    \centering
    \includegraphics[width=85mm]{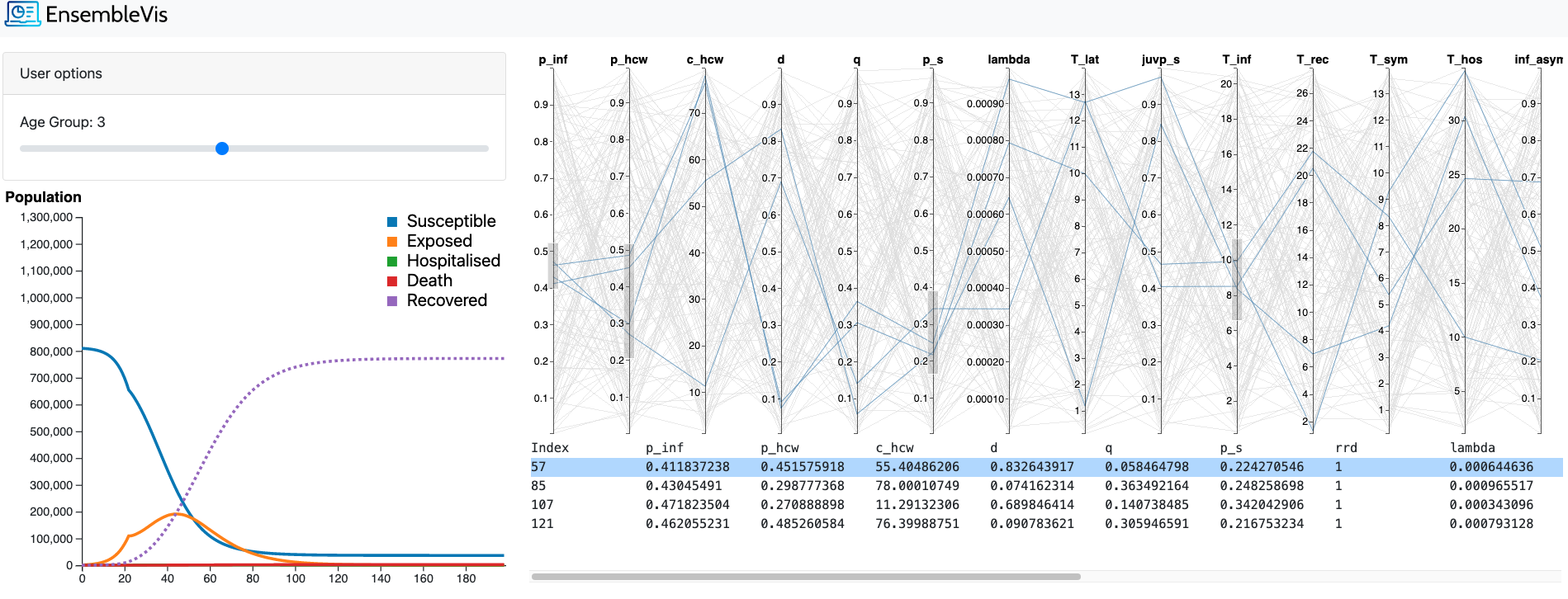}
    \caption{A prototype system for analysing and visualising ensemble data. The parallel coordinate plot (right) allows one to filter input parameters, select a subset of time series to be aggregated, and display the aggregated curves (left).}
    \label{fig:ModellingD}
    \vspace{-5mm}
\end{figure}

\vspace{-2mm}
\subsection{Team M$_D$: Supporting Inference and Model Assessment}
\label{sec:ModellingD}
One group of SCRC modelling scientists have been focused on quantifying epidemic characteristics, effect of intervention, and model performance.
From the onset, the VIS volunteers for supporting these modelling activities anticipated the use of ensemble and uncertainty visualisation.
During an initial collaboration meeting, domain experts also confirmed the need for visualising the sensitivity of model parameters.
The domain experts were working on their models in parallel, thus VIS volunteers were not able to obtain multi-run simulation data in the early months of the collaboration.
VIS Team $\mathbf{M}_D$ took initiatives to study two COVID-19 models in the public domain, attempting to generate multi-run simulation data.
Before this attempt could yield useful output, one SCRC model, ABC-smc \cite{EERAModel:2020:git} produced multi-run simulation data for uncertainty visualisation and parameter space analysis.
By attending modelling scientists' meetings, the team were able to observe the interactions between the perspectives of modelling and uncertainty quantification, and established a set of requirements.

Similar to most multi-run simulation problems, VIS needed to support the analysis of many sets of model parameters and outputs. In this case, each dataset consists of some 200 time series and their corresponding parameter sets.
One obvious requirement is to visualise the uncertainty featured in the set of time series.
As this is a common requirement for all models with time series outputs, we passed the requirement to the generic support team (see Figure \ref{fig:UI}).
The team focused on the more complex tasks, i.e., (i) to identify input/output relationships, (ii) to determine key curve features such as maximum or largest slope, and (iii) to compare outputs from a number of different model runs. Immediately after the requirements analysis, we started to develop a VIS system iteratively, with increasing facilities for analytics, visualisation, and interaction.   

Figure~\ref{fig:ModellingD} shows the current prototype after several iterations.
We used the design approach of coordinated multiple views~\cite{boukhelifa2003coordination},
the visual analytics approach for computing a set of curve features~\cite{tam2011visualization},
a parallel coordinate plot~\cite{Inselberg:1990} for viewing and filtering multi-dimensional
parameter sets and curve features, and aggregated curves summarising the outputs of the selected parameter sets. We are in the process of introducing new facilities, such as slicing the multi-dimensional parameter space~\cite{torsney2017sliceplorer}, using functional box
plots\cite{whitaker2013contour,mirzargar2014curve} to summarize many curves, and the contribution-to-the sample-mean plot~\cite{bolado2009contribution} to show the sensitivity of outputs to input parameters. 

\begin{figure*}[t]
	\centering
	\includegraphics[width=170mm]{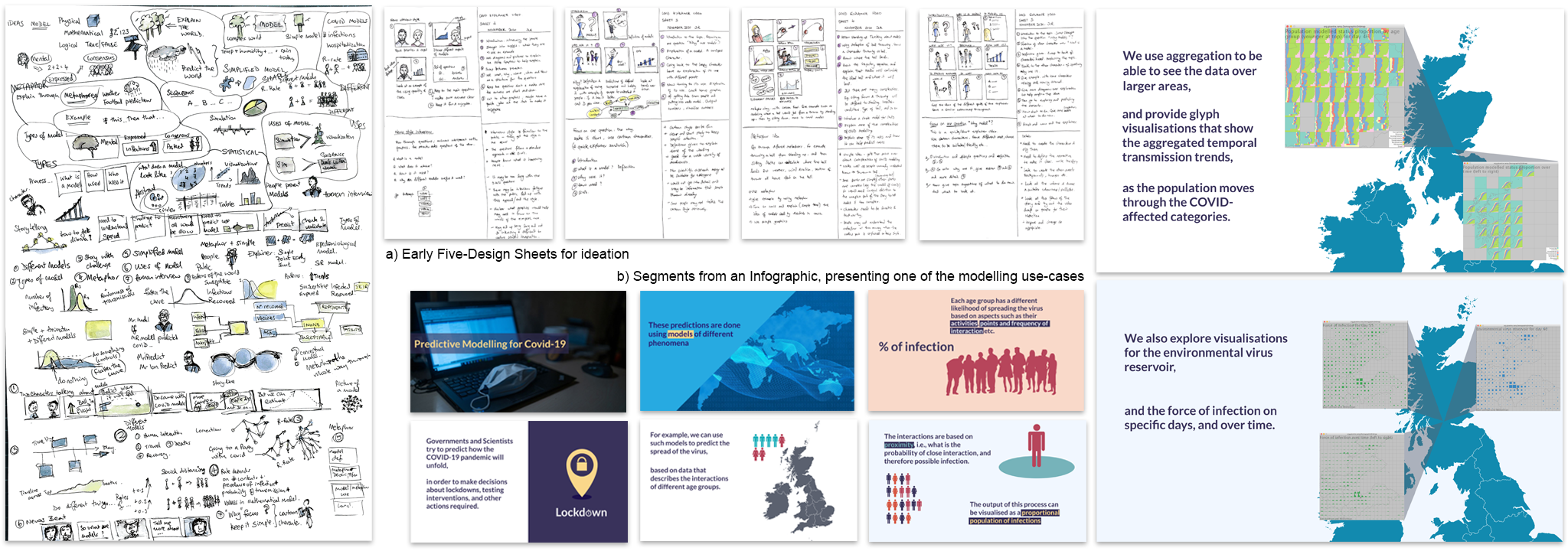}
	\caption{Storytelling visualisation creation process. (a) Initial concepts and ideas were explored using a combination of the five-design sheets~\cite{RobertsFDS_2016} method. (b) Following the progress of the modelling activities, these were transformed to animated presentations and infographics, which incorporated visualisations from other teams (in this example, from Team $\mathbf{M}_A$).}
	\label{fig:Dissemination}
	\vspace{-4mm}
\end{figure*}

\section{Disseminative Visualisation}
\label{sec:Dissemination}
While most VIS volunteers were distributed among the aforementioned six teams, we also created a small team for \emph{disseminative visualisation} since public information dissemination was the original overall requirement.
One epidemiology researcher with interest in data visualisation also joined the team.
The team explored the prevailing approaches, in the UK and internationally, in public-facing visualisations related to the pandemic. This ranged from those produced by a number of governments (e.g., the four home nations in the UK), organisations (e.g., WHO, UK ONS), universities (e.g., Johns Hopkins dashboards), media outlets (e.g., FT Coronavirus tracker), and non-commercial web services (e.g., Worldometers).

The team concluded that we should complement, but not duplicate, the existing effort, and defined our goal as to inform the public about activities of SCRC through storytelling visualisation.
We identified the following requirements: (i) to maintain scientific rigour, (ii) to retain scientific language, and (iii) to abstract visualisation output.
For example, one of our initial designs used  football  result prediction and weather forecasting metaphors.
Our rigorous consideration indicated that the former might be associated with gambling, whereas the latter could be perceived as inaccurate, and they could have negative connotations to how epidemiological modelling would be perceived.

Figure~\ref{fig:Dissemination} illustrates our process for creating  storytelling visualisation.
It started with an ideation phase where we elaborated preliminary concepts and ideas. This was done using a combination of the five-design sheets~\cite{RobertsFDS_2016} methodology (Figure~\ref{fig:Dissemination}(a)), animated PowerPoint mock-ups, and web-based prototyping. These sparked off other explanatory forms, e.g., infographics and slide packs, such as in Figure~\ref{fig:Dissemination}(b), which were used to describe each epidemiological model and the corresponding visualisations. As it would be difficult to apply a unified narrative to every model, we opted for allowing each story to be developed independently.

Current storytelling visualisations have been implemented as: (a) web-based presentations using the Reveal.js framework, with SVG-based animations and the potential for directly feeding into them visualisations created by other teams, (b) as video outputs of animated presentations, and (c) as infographics, created using graphics editors and creative design tools. We are in the process of creating a public web server for hosting these storytelling visualisations. 

\section{Reflections and Recommendations}
\label{sec:Reflection}

In this section, we reflect on our experience of developing VIS capacities for emergency response, and translate our reflections to a set of recommendations as a step towards a new methodology. 

\noindent\textbf{Reflection on general perception of VIS as a dissemination tool.}
Expert users often see visualisation as ``for informing others'' rather than ``for helping myself''. This can be a big stumbling block during requirement analysis. As illustrated in Figure \ref{fig:Sketch}, Dr. Reeve’s response during the first meeting helped overcome this stumbling block, shortening the delay in requirements analysis by months.
Meanwhile, to the disseminative visualisation team, creating such visualisation has not been an easy journey, especially without the advice from an expert on public engagement.  

\noindent \textbf{Reflection on requirements analysis.} During the pandemic, domain experts were extremely busy. Different teams did not follow the same formula for requirement analysis. Modelling support teams $\mathbf{M}_B$ and $\mathbf{M}_C$ followed the recommended method for user-centred requirement analysis, and benefited from frequent engagement. The generic support team and team $\mathbf{M}_A$ identified urgent requirements quickly and began their development without much delay. The analytical support team, team $\mathbf{M}_D$, and disseminative visualisation team had to use their knowledge to anticipate and analyse the potential requirements. In emergency responses, all of these are valid methods.
Several teams had positive experience in using quickly-produced visualisations to stimulate requirement analysis.
Team $\mathbf{M}_B$ made good use of several communication mechanisms, including searchable threaded chat streams, structured feedback, and design exposition. A few VIS volunteers also found it rewarding to attend domain experts' meetings that at first appeared irrelevant.

\noindent \textbf{Reflection on team organization.}
The categorisation of visualisation tasks based on the complexity of the search space of the possible solutions is relatively new \cite{Chen:2016:TVCG}. It informs us that model optimisation is an NP process in general, and it requires all three levels of visualisation, i.e., observational, analytical, and model-developmental visualisation, which correspond to solution spaces of complexity $O(n)$, $O(n^k)$, and NP. The complexity is likely to impinge on the effort for identifying VIS requirements. Hence, having a VIS team working with each modelling team was necessary for establishing such understanding.
All teams quickly identified and addressed the observational requirements related to individual models, and some have started to address the requirements for model analysis and model optimisation.
Meanwhile, the generic support team progressed to the development stage quickly because of not only the necessity but also the less complex search space.

\noindent \textbf{Reflection on VIS resources.}
Using volunteer effort is not an ideal solution for emergency responses. It would be more efficient if we could utilise an existing technical and knowledge infrastructure for such an emergency response, if such an infrastructure had existed for other operations and had an advanced VIS server and a team of VIS developers who were knowledgeable about different levels of visualisation tasks. Our volunteering effort was a make-shift solution, which benefited strongly from the academic knowledge infrastructure in the UK. Its progress could be more rapid if there were more development resources.
The organisation of VIS volunteers partly reflects the need to concentrate most development resources in the generic support team.
Nevertheless, the outcomes delivered by the VIS volunteers within the past six months without any funding are unprecedented. This demonstrates the importance of VIS as well as volunteering effort in emergency responses.

\noindent\textbf{Recommendation.} Our approaches, experience, and reflections may be translated to the following recommendations for future VIS applications in emergency responses:
\begin{itemize}
    \item In December 2014, US President Barack Obama spoke to the National Institutes of Health (USA): ``There may and likely will come a time in which we have both an airborne disease that is deadly. And in order for us to deal with that effectively, we have to put in place an infrastructure'' \cite{Obama:2014:web}. VIS should be part of such an infrastructure, and the ``readiness'' of VIS technical and knowledge infrastructures will make a difference.
    \item While the agile principle \cite{10Munzner} fits well with VIS development for supporting emergency responses, one should be open-minded about different approaches. The diverse approaches taken by different VIS teams in the RAMP VIS effort indicated that standard practice might not always be applicable.
    \item VIS development in emergency responses can benefit tremendously from the existing VIS knowledge, in the form of theories, methodologies, literature, and personal experience.
    The VIS community should improve its ``readiness'' by advancing
    abstract VIS knowledge in the form of theories and methodologies.
\end{itemize}

\section{Conclusions}

In this paper, we have reported the work carried out by a group of VIS volunteers to support modelling scientists and epidemiologists in combating COVID-19. Our approaches to the challenges that we have encountered are rare and valuable contributions to the first step towards a methodology for developing and providing VIS capacity to support emergency response. The UK Research and Innovation recently awarded funding to the group, transforming the volunteering effort to a more structured VIS operation in 2021.

\section*{Acknowledgement}
The authors would like to express their gratitude to the
SCRC management team for its leadership, all SCRC members for their selfless efforts, the UK 
Science and Technology Facilities Council (STFC) for providing hardware to SCRC and RAMP VIS, and the STFC for providing and managing the SCRC hardware platform.

In particular, we would like to thank
Dr. Rita Borgo who led the activity described in Section \ref{sec:ModellingD} but could not be a co-author due to her co-chair role in EuroVis 2021.
We thank other VIS volunteers, including Professor Nigel W. John, Dr. Helen C. Purchase, Dr. Stella Mazeri, Dr. William Teahan, Benjamin Nash, Tianci Wen, Dr. Dylan Rees, Elif E. Firat, and James Scott-Brown for their contributions to various VIS activities.
In addition, we acknowledge that some VIS activities have received support from
Dr. Alessio Arleo, Peter Butcher, Cameron Gray, Jeewan Hyongju, and Edison Mataj.

We would like to express our appreciation for
Andrew Lahiff for setting up and managing the RAMP VIS VMs, 
Sonia Michell for offering innumerable pieces of advice on data products,
and Dr. Iain McKendrick for leading the discussion on authorship guidelines. 

Last but not least, our big thanks go to the modelling scientists and epidemiologists who worked with the VIS teams closely and provided valuable expert advice as well as datasets to be visualized.
They include Dr. Jess Enright, Dr. Claire Harris, Professor Louise Matthews, Dr. Glenn Marion, Dr. Sibylle Mohr, and Dr. Thibaud Porphyre.

\bibliographystyle{eg-alpha-doi}  
\bibliography{RAMPVIS,RAMPVIS2}        


\end{document}